\documentstyle[12pt]{article}

\advance \textheight by \topskip
\makeatletter
\@addtoreset{equation}{section}

\makeatother

\begin {document}

\begin{titlepage}
\hbox to \hsize{\hfil {\tt BONN-TH-98-07}}
\hbox to \hsize{\hfil {\tt February 1998}}
\vfill
\large \bf
\begin{center}
Pseudoclassical mechanics and \\
hidden symmetries of 3d particle models
\end{center}
\vskip1cm
\normalsize
\begin{center}
Khazret S. Nirov{\footnote{E--mail: 
nirov@dirac.physik.uni-bonn.de\hskip1cm
(Alexander von Humboldt fellow)}${}^{,}${}\footnote{On leave 
from the {\sl Institute for Nuclear Research of the Russian 
Academy of Sciences}}}\\
{\small \it Physikalisches Institut, Universit\"at Bonn, 
Nussallee 12, D-53115 Bonn, Germany} 
\end{center}
\vskip2cm
\begin{abstract}
\noindent
We discuss hidden symmetries of three-dimensional field 
configurations revealed at the one-particle level by the 
use of pseudoclassical particle models. We argue that at
the quantum field theory level, these can be naturally 
explained in terms of manifest symmetries of the reduced 
phase space Hamiltonian of the corresponding field systems.
\end{abstract}
\vskip4cm
\begin{center}
{\sl Talk presented at the Meeting on}\\ 
{\it Quantum Aspects of Gauge Theories, 
Supersymmetry and Unification}\\ 
{\sl Neuch\^atel, Switzerland, 18--23 September 1997} 
\end{center}
\vfill
\end{titlepage}

\hspace{2cm}
\section{Introduction}

Pseudoclassical mechanics has been basically understood in terms 
of the principles formulated by J.L. Martin (1959) \cite{JLM}, 
F.A. Berezin and M.S. Marinov (1975-77) \cite{BM}, and R. Casalbuoni
(1976) \cite{RC}. The underlying idea of refs. \cite{JLM,BM,RC} was 
to consider elements of a Grassmann algebra as classical dynamical 
variables (functions on the phase space), so generalizing the concept 
of classical mechanics to a purely algebraic construction of the ring 
and the graded Lie algebra, in order to describe {\it spin dynamics}. 
Such a formalism was implemented also by A. Barducci, R. Casalbuoni 
and L. Lusanna, and L. Brink, S. Deser, B. Zumino, P. DiVecchia and 
P. Howe \cite{PM2}, when constructing various models of supersymmetric 
spinning particles. 

There, as the starting point, the well-known covariant supersymmetric 
differential 1-form was reconsidered. It was observed that this object 
can be realized in different ways. Indeed, instead of considering the 
superspace parameterized by the usual space-time coordinates and pairs 
of Weyl spinors, one can introduce real anticommuting variables, 
Grassmannian odd (pseudo)vectors and (pseudo)scalars, together 
with the even classical coordinates forming a set of configuration 
space variables. Upon quantization, these new variables turn to be 
the generators of real Clifford algebras with corresponding spin 
indices, but not the Fermi variables as it was formerly, so that 
no contradiction with the spin-statistics relation emerges. 

Having all this at hand, one can construct various superparticle 
(pseudoclassical) Lagrangians, thus explicitly describing spin 
degrees of freedom of the corresponding quantum-mechanical and
field systems. Further, if the latter possess some dynamical 
(hidden) symmetries, for which the spin dynamics is responsible, 
then the formalism of pseudoclassical mechanics certainly provides 
an appropriate basis for revealing such properties and understanding 
their nature.

Exactly this task was in the program of refs. \cite{GPS,NP1}, 
where hidden symmetries of three-dimensional field configurations 
were revealed by means of the corresponding pseudoclassical models.
So, the spin dynamics of the $P$,$T$-invariant systems of planar
massive fermions \cite{GPS} and topologically massive U(1) gauge
fields \cite{NP1} have been described by the Lagrangians $L_F$ 
\cite{CPV} and $L_T$ \cite{NP1,NP2}, respectively with
\begin{eqnarray}
L_F &=& \frac{1}{2e} \left(\dot{x}_\mu - i v
\varepsilon_{\mu\nu\lambda}\xi^\nu \xi^\lambda \right)^2
- \frac{1}{2} e m^2 + 2 i qmv\theta_1\theta_2 - \frac{i}{2}
\xi^\mu \dot\xi_\mu + \frac{i}{2}\theta_a\dot\theta_a,
\label{LF} \\
L_T &=& \frac{1}{2e} \left(\dot{x}_\mu - \frac{i}{2} v
\varepsilon_{\mu\nu\lambda}\xi^\nu_a \xi^\lambda_a \right)^2
- \frac{1}{2} e m^2 - i qmv\xi^\mu_1 \xi_{2\mu} + \frac{i}{2}
\xi^\mu_a \dot\xi_{a\mu}.
\label{LT}
\end{eqnarray}
The configuration spaces of the systems were thus represented 
by the sets of even, $x_\mu$, $e$, $v$, and odd, $\xi^\mu$,
$\theta_a$, $\xi^\mu_a$, variables, $\mu=0,1,2$, $a=1,2$. 
In this, $x_\mu$ are the space-time coordinates of the particles, 
$e$ and $v$ make sense as Lagrange multipliers, with $e$ representing 
the world-line metrics, whereas the odd variables $\xi^\mu$,
$\theta_a$ and $\xi^\mu_a$ are incorporated for the spin degrees 
of freedom of the three-dimensional field theories expected as 
quantum counterparts of the pseudoclassical models; $q$ denotes 
a real $c$-number parameter, hereafter called the model parameter; 
the metric and the totally antisymmetric tensors of the 3d Minkowski 
space-time are fixed by the conditions $\eta_{\mu\nu} = diag(-1,+1,+1)$ 
and $\varepsilon^{012} = 1$.

The models (\ref{LF}) and (\ref{LT}) have rich sets of symmetries -- 
discrete, gauge and continuous global. It was observed \cite{GPS,NP1}
that $L_F$ and $L_T$ hold parity and time-reversal invariance (discrete
symmetries) for any value of the model parameter. However, the cases of
$q = -1, 0, +1$ in (\ref{LF}) and $q = -2, 0, +2$ in (\ref{LT}) turned
out to be special both at the classical and the quantum levels of the
theories. The case of $q = 0$ is dynamically degenerated, with some
of the spin variables being trivial integrals of motion, and this
value of the model parameter is excluded at the quantum level. 
As well, the values $q = -1, +1$ and $q =-2, +2$ are separated 
from the point of view of dynamics of (\ref{LF}) and (\ref{LT}), 
respectively. However, in the latter cases, the classical systems
get maximal global symmetries, with sets of generators including
nontrivial integrals of motion which exist only at these special
values of the model parameter. Exactly the same values reveal
themselves at the quantum level too; but quantum mechanically
they are singled out not only by the requirement for the
continuous global symmetries to be maximal. Indeed, only 
for these special values of the model parameter the discrete 
symmetries of the initial classical systems are conserved upon
quantization. We see that if a system has a quantized parameter,
then its special discrete values may in principle reveal themselves
just at the classical level, from the point of view of dynamics.
This interesting phenomenon, called in ref. \cite{NP2}
{\it the classical quantization}, implies some hidden relationships 
between continuous and discrete symmetries. Note that (\ref{LF}) and 
(\ref{LT}) belong to the class of gauge systems having quadratic in 
Grassmann variables nilpotent constraints of the form \cite{NP2}
\[
\left( A_{ij}\delta^{ab} + q B_{ij}\epsilon^{ab} \right)
\zeta^i_a \zeta^j_b \approx 0,
\]
where $A_{ij}$ and $B_{ij}$ being functions of even variables
fulfil the relations 
$A_{ij} = - A_{ji}$, $B_{ij} = B_{ji}$,
$\delta^{ab}$ is the Kronecker symbol, $a,b = 1,2$,
$\epsilon^{ab} = -\epsilon^{ba}$.
It can be shown that in this general case, the special values of 
$q$ are singled out by the following quantization condition:
\[
\{ q = q_k \;\vert\; {\rm det}\; \Vert 2ik A_{ij} 
+ q_k ( k B_{ij} - \frac{1}{2} B \eta_{ij} \Vert = 0 \},
\qquad
k = 1,\ldots,K,
\]
$B = \eta^{ij} B_{ij}$ and $\eta_{ij}$ is regarded as the metric 
tensor of the $K$-dimensional linear space spanned by the indices 
$i,j = 1,2,\ldots,K.$ The quantization condition for the model 
parameter means actually consistency of classical and quantum 
dynamics. 

Quantization of the systems with the Lagrangians $L_F$ and $L_T$ 
has led to three-dimensional parity and time-reversal conserving 
systems of massive fermions and Chern-Simons U(1) gauge fields, 
respectively \cite{GPS,NP1}. When analyzing algebras of the 
integrals of motion of the pseudoclassical models, hidden U(1,1) 
symmetry and S(2,1) supersymmetry of the corresponding field 
configurations were elucidated \cite{GPS,NP1}. It has also been 
shown that these field systems realize irreducible representations 
of a non-standard super-extension of the (2+1)-dimensional 
Poincar\'e group, ${\rm ISO}(2,1 \vert 2,1)$, labelled by 
the zero eigenvalue of the corresponding superspin operators. 

In this talk, we shall discuss dynamical U(1,1) symmetry of the 
$P$,$T$-invariant system of topologically massive gauge fields,
related to (\ref{LT}), and show that this hidden symmetry can 
be naturally understood in terms of manifest symmetries of the 
corresponding reduced phase space Hamiltonian. Similar analysis 
for the double fermion system can analogously be provided. With 
this investigation, we start solving the problem of generalization 
of the hidden symmetries, revealed at the one-particle level 
\cite{GPS,NP1}, onto the level of quantum field theory. 

In the section 2, we recall the structure of the field system
in question and stress on similarity to the string field theory
construction; then, in the section 3, we carry out the Hamiltonian
description of the system. The paper is concluded by discussion
of some related problems to be further investigated.

\section{Constructing the $P$,$T$-invariant field system}

In order to find the field system corresponding to the 
pseudoclassical model (\ref{LT}), its general quantum 
state $\Psi(x)$ was realized over the vacuum as an expansion 
into the complete set of eigenvectors of the fermion number 
operator \cite{NP1}. The coefficients of this expansion, 
initially supposed to be square-integrable functions of the 
space-time coordinates, turned in fact to be belonging to the 
Schwartz space, which is a rigged Hilbert space \cite{BLOT}. 
The latter is due to the mass-shell, $\phi \approx 0$, and 
quadratic in spin variables nilpotent, $\chi \approx 0$, 
constraints following from the Lagrangian $L_T$. The quantum 
counterparts of these first class constraints were used to 
single out the physical subspace of the theory, 
$\widehat\phi \Psi = \widehat\chi \Psi = 0$. 
As the consistent solution to the quantum constraints, 
the {\it doublet} of vector fields 
${\cal F}^\mu_+$, ${\cal F}^\mu_-$ 
satisfying the linear differential equations
\begin{equation}
{\cal L}^\pm_{\mu\nu} {\cal F}^\nu_\pm = 0, 
\qquad
{\cal L}^\pm_{\mu\nu} \equiv 
\varepsilon_{\mu\nu\lambda} \partial^\lambda 
\pm m \eta_{\mu\nu},
\label{BE} 
\end{equation}
was obtained. This is the subspace of the total state
space connected with the special values $|q| = 2$ of
the model parameter. It is easy to see that, due to the 
basic equations (\ref{BE}), the fields ${\cal F}^\mu_\pm$
obey also the Klein-Gordon equation 
$(-\partial^2 + m^2) {\cal F}^\mu_\pm = 0$ 
and the transversality condition
$\partial_\mu {\cal F}^\mu_\pm = 0$. 
Therefore, the physical subspace of the theory is described
by the vector fields ${\cal F}^\mu_\pm$, respectively
carrying massive irreducible representation of the spin 
$s = \mp 1$ of the 3d Poincar\'e group. 

To construct the action functional, that would reproduce 
the field equations (\ref{BE}) by the variational principle, 
the average value of the constraint operator $\widehat\chi$ 
over the general state was considered,
$\langle \widehat\chi \rangle 
= \Psi^\dagger(x) \widehat\chi \Psi(x)$.
It was however observed that the metric on the state space 
was indefinite in the doublet 
$\Phi = { {\cal F}_+ \choose {\cal F}_-}$.  
Actually, restricting the scalar product
$\langle\, , \rangle$ onto the physical subspace, one gets
$\langle \Psi , \Psi \rangle = \bar{\Phi} \Phi$,
$\bar\Phi = ({\cal F}_+ , - {\cal F}_- )$.
And so, to have the norm of the state vectors defined from
a positive-definite scalar product and give the physical
states with spins $+1$ and $-1$ equal treatment, the scalar
product was modified to 
$\langle\langle\, , \rangle\rangle = \bar{\Phi}{\sigma_3}{\Phi}$.
For the constraint operator $\widehat\chi$ this gave the
following average value restricted onto the physical subspace:
\begin{equation}
\langle\langle \widehat\chi \rangle\rangle 
= {\varepsilon^\alpha}_{\mu\beta} \left( 
{\cal F}_{+\alpha} \partial^\mu {\cal F}^\beta_+ 
+ {\cal F}_{-\alpha} \partial^\mu {\cal F}^\beta_- 
\right) + m \left( {\cal F}_{+\gamma} {\cal F}^\gamma_+ 
- {\cal F}_{-\gamma} {\cal F}^\gamma_- \right). 
\label{chi}
\end{equation}
The space-time integral of this quantity has finally resulted 
in the desirable action functional 
\begin{equation}
{\cal A} = \int d^3x \langle\langle \widehat\chi \rangle\rangle
= \int d^3x \left( 
{\cal F}^\mu_+ {\cal L}^+_{\mu\nu} {\cal F}^\nu_+ 
+ {\cal F}^\mu_- {\cal L}^-_{\mu\nu} {\cal F}^\nu_- \right) 
\label{A}
\end{equation}
of the parity and time-reversal conserving system of Chern-Simons
U(1) gauge fields \cite{CS}, given in terms of a self-dual free 
massive field theory \cite{self}. 

Dynamical (super)symmetries of the field theory (\ref{A}) are
generated by the average values of the quantum counterpart of
the integrals of motion of the system (\ref{LT}) with $|q| = 2$.
In this, the corresponding quantum mechanical nilpotent operators
\cite{NP1} realize mutual transformation of the physical states of 
spins $+1$ and $-1$. It is crucial that these physical operators 
turned out to be Hermitian exactly with respect to the scalar 
product $\langle\langle\, , \rangle\rangle$. 

The procedure we have implemented is actually reminiscent of that 
suggested by Siegel \cite{Siegel} for constructing a string field 
theory and subsequently developed by Witten \cite{Witten}. 
There, an object of the form 
$A = \int d\mu \langle\Psi\vert\Omega\vert\Psi\rangle$,
with a BRST operator $\Omega$ singling out physical states 
and $d\mu$ being an integration measure, was treated as a 
string field theory action. Also, a scalar product 
$\langle\vert\vert\rangle$ was proposed to ensure hermiticity 
for the BRST operator. The underlying idea was taken from the 
observation that the functional $A$ is extremal on the physical
subspace: the variational principle applied to the ``action'' 
$A$ reproduces ``quantum equations of motion'' encoded in the 
BRST condition $\Omega\vert\Psi\rangle = 0$, and besides, it keeps 
symmetries of the initial first-quantized theory. Obviously, here 
we dealt with an analogous construction, while having the constraint 
operator $\widehat\chi$ instead of the BRST-charge $\Omega$ and the 
finite-mode decomposition of the general state vector $\Psi(x)$ where 
the expansion coefficients were, in particular, topologically massive 
U(1) gauge fields. As well as in the string field theory 
\cite{Siegel,Witten}, the action ${\cal A}$ detains all 
symmetries of its first-quantized counterpart \cite{NP1}.

\section{Hamiltonian on the reduced phase space}

To construct the Hamiltonian description of the field theory 
(\ref{A}), let us introduce, as usual, the phase space with 
the generalized momenta ${\cal P}^\mu_\epsilon(\vec{x},t)$ 
canonically conjugate to the generalized coordinates 
${\cal F}^\mu_\epsilon(\vec{x},t)$, $\epsilon = +,-$, 
\begin{equation}
\{ {\cal F}^\mu_\epsilon(\vec{x},t) , 
{\cal P}^\nu_{\epsilon^\prime}(\vec{y},t) \} = 
\delta_{\epsilon\epsilon^\prime} \eta^{\mu\nu}
\delta(\vec{x} - \vec{y}).
\label{pois}
\end{equation}
There are primary constraints in the system,
\begin{equation}
{\cal P}^0_\epsilon \approx 0, \qquad
{\cal P}^i_\epsilon 
+ \varepsilon^{0ij} {\cal F}^j_\epsilon \approx 0,
\label{prim}
\end{equation}
and hence, the total Hamiltonian 
$H_T(t) = \int d^2\vec{x}\; h_T(\vec{x},t)$
contains Lagrange multipliers
$\lambda^0_\epsilon$ and $\lambda^i_\epsilon$; we have
\[
h_T = h + \sum_{\epsilon = +,-} 
\lambda^0_\epsilon {\cal P}^0_\epsilon + \lambda^i_\epsilon 
\left( {\cal P}^i_\epsilon 
+ \varepsilon^{0ij} {\cal F}^j_\epsilon \right),
\]
where the density of the canonical Hamiltonian $h(\vec{x},t)$ 
is given by the expression
\[
h = \sum_{\epsilon = +,-} \varepsilon^{0ij} \left(
{\cal F}^j_\epsilon \partial_i {\cal F}^0_\epsilon
- {\cal F}^0_\epsilon \partial_i {\cal F}^j_\epsilon \right)
+ \epsilon m \left( {\cal F}^0_\epsilon {\cal F}^0_\epsilon 
- {\cal F}^i_\epsilon {\cal F}^i_\epsilon \right).
\]
Under subsequent application of the Dirac-Bergmann formalism 
\cite{DB}, secondary constraints arise in the system,
\begin{equation}
\varepsilon^{0ij} \partial_i {\cal F}^j_\epsilon 
- \epsilon m {\cal F}^0_\epsilon \approx 0, 
\label{sec}
\end{equation}
and the Lagrange multipliers get fixed,
$\lambda^0_\epsilon = - \partial_i {\cal F}^i_\epsilon$,
$\lambda^i_\epsilon = - \partial^i {\cal F}^0_\epsilon 
+ \epsilon m \varepsilon^{0ij} {\cal F}^j_\epsilon$.
The complete set of the constraints (\ref{prim}), (\ref{sec}) is
of the second class. We see that the initial phase space can be
reduced to the corresponding constraint surface. Restricting the
Poisson brackets (\ref{pois}) onto this reduced phase space, one
obtains the related Dirac brackets. The physical degrees of freedom,
i.e. the points of the reduced phase space, may now be described
by the set of the field variables ${\cal F}^i_\epsilon(\vec{x},t)$,
$i=1,2$, $\epsilon=+,-$, having the Dirac brackets
\begin{equation}
\{ {\cal F}^i_\epsilon(\vec{x},t) , 
{\cal F}^j_{\epsilon^\prime}(\vec{y},t) \}^* = 
\frac{1}{2} \delta_{\epsilon\epsilon^\prime} \varepsilon^{0ij} 
\delta(\vec{x} - \vec{y})
\label{Db}
\end{equation}
and satisfying the equations of motion
\[
\dot{\cal F}^i_\epsilon = -\epsilon \varepsilon^{0ij}\,
\Delta_{jk}\, {\cal F}^k_\epsilon, 
\qquad 
\Delta_{ij} = m \delta_{ij} - \frac{1}{m} 
\varepsilon_{0ik} \partial^k \varepsilon_{0jl} \partial^l.
\]
It is worthwhile noting that one has also the equations
$\dot{\cal F}^0_\epsilon = - \partial_i {\cal F}^i_\epsilon$
confirming the above mentioned transversality condition. 
In terms of these variables, converted into the representation
\begin{equation}
{\cal F}^i_\epsilon(\vec{x},t) = \frac{1}{2\pi} \int d^2\vec{p}\;
{\rm e}^{i\vec{p}\vec{x}}\, \tilde{\cal F}^i_\epsilon(\vec{p},t),
\qquad
\overline{\tilde{\cal F}}{}^i_\epsilon(\vec{p},t) 
= \tilde{\cal F}^i_\epsilon(-\vec{p},t), 
\label{arep}
\end{equation}
the Hamiltonian of the system on the reduced phase space takes 
the form
\begin{equation}
H^* = - \int d^2\vec{p} \sum_{\epsilon=+,-} \epsilon \;
\tilde{\cal F}^i_\epsilon\, \tilde\Delta_{ij}\, 
\tilde{\cal F}^j_\epsilon,
\qquad
\tilde\Delta_{ij} \equiv m \delta_{ij} 
+ \frac{1}{m} \varepsilon_{0ik} p^k 
\varepsilon_{0jl} p^l.
\label{hamr}
\end{equation}
Seeing that the operator $\tilde\Delta_{ij}$ allows for the
representation
\[
\tilde\Delta_{ij} = \omega_{ik} {\omega^k}_j, \qquad
\omega_{ij} \equiv \sqrt{m} \delta_{ij} + \frac{1}{\sqrt{m}}
\frac{\varepsilon_{0ik}p^k\varepsilon_{0jl}p^l}{\sqrt{m^2+\vec{p}^2}+m},
\]
so that the new fields 
$u^\epsilon_i = \omega_{ij} \tilde{\cal F}^j_\epsilon$ 
are well-defined, the Hamiltonian $H^*$ can be given another 
convenient form
\[
H^* = \int d^2\vec{p} \left( u^-_i u^-_i - u^+_i u^+_i \right).
\]
Finally, passing on to the complex variables
$z_\epsilon = u^\epsilon_1 + i u^\epsilon_2$ 
with the canonical brackets
\[
\{ z_\epsilon(p) , \bar{z}_{\epsilon^\prime}(q) \}^*
= - i |p^0| \delta_{\epsilon\epsilon^\prime}
\delta(\vec{p} + \vec{q}), 
\qquad
|p^0| \equiv \sqrt{m^2 + \vec{p}^2},
\]
we get for the reduced phase space Hamiltonian the expression
\begin{equation}
H^* = \int d^2\vec{p} \left( |z_-|^2 - |z_+|^2 \right).
\label{H*}
\end{equation}
The latter is exactly the well-known \cite{BR} defining quadratic 
form for the symmetry group U(1,1) with the non-compact 
group manifold ${\cal M}({\rm U}(1,1)) \cong S^1 \times R^2$. 
Note that if one rescales the complex variables $z_\pm$ as
${|p^0|^{1/2}} z_\epsilon \rightarrow z_\epsilon$, they become 
the classical counterparts of the oscillator annihilation-creation 
operators obeying the standard commutation relations
\begin{equation}
[\widehat{z}_\epsilon(p),\widehat{\bar z}{}_{\epsilon^\prime}(q)]
= \delta_{\epsilon\epsilon^\prime} \delta(\vec{p} + \vec{q}).
\label{1}
\end{equation}
Let us introduce a positive-definite scalar product $( \, , )$
on the corresponding state space and denote the respective Hermitian 
conjugation by the dagger $\dagger$. Then the operators 
$\widehat{z}_\epsilon$ and $\widehat{\bar z}{}_\epsilon$ are mutually 
conjugate with respect to the scalar product $( \,, )$. We have
\begin{equation}
{(\widehat{z}_\epsilon)}^\dagger = \widehat{\bar z}{}_\epsilon.
\label{2}
\end{equation}
The explicit forms of the dynamical symmetry group generators can 
easily be found (see e.g. ref. \cite{BR}). They are simply the 
conserved charges of the field theory with the Hamiltonian being 
the quantum counterpart of $H^*$. Their densities are given by the
expressions
\begin{equation}
u = \frac{1}{2}\left( \widehat{\bar z}{}_-\widehat{z}_- 
- \widehat{\bar z}{}_+\widehat{z}_+ \right), \quad
q_0 = \frac{1}{2}\left( \widehat{\bar z}{}_-\widehat{z}_- 
+ \widehat{\bar z}{}_+\widehat{z}_+ + 1 \right), \quad
q_+ = \widehat{z}_+\widehat{z}_-, \quad
q_- = \widehat{\bar z}{}_-\widehat{\bar z}{}_+ .
\end{equation}
In this, $u$ is the density of the U(1) generator, 
while the rest fulfil the usual $su(1,1)$ algebra 
\[
[ q_+ , q_- ] = 2 q_0, \qquad
[ q_\pm , q_0 ] = \pm q_\pm.
\]
The operators $u$ and $q_0$ are Hermitian, and $q_+$ and $q_-$
are mutually conjugate with respect to the scalar product $(\,,)$,
${(u)}^\dagger = u$, ${(q_0)}^\dagger = q_0$, 
${(q_\pm)}^\dagger = q_\mp$.
We have thus reproduced the dynamical 
${\rm U}(1,1) = {\rm U}(1)\times{\rm SU}(1,1)$ symmetry of the 
$P$,$T$-invariant system of topologically massive vector fields 
at {\it the quantum field theory level}. Certainly, one could 
gain the same ends while using not the auxiliary representation
(\ref{arep}), but the ordinary decomposition of the fields
${\cal F}^i_\epsilon$ over annihilation and creation operators.

Instead of dealing with the action ${\cal A}$, one might consider
another action functional
\begin{equation}
{\cal A}^\prime = \int d^3x \left( 
- {\cal F}^\mu_+ {\cal L}^+_{\mu\nu} {\cal F}^\nu_+ 
+ {\cal F}^\mu_- {\cal L}^-_{\mu\nu} {\cal F}^\nu_- \right) 
\label{A'}
\end{equation}
leading to {\it the same equations} (\ref{BE}) for the Chern-Simons 
vector fields. Constructing the reduced phase space Hamiltonian reads:
\begin{equation}
H^{\prime *} = \int d^2\vec{p} \sum_{\epsilon=+,-}
\tilde{\cal F}^i_\epsilon\, \tilde\Delta_{ij}\, 
\tilde{\cal F}^j_\epsilon = 
\int d^2\vec{p} \left( |z_-|^2 + |z_+|^2 \right).
\label{H'}
\end{equation}
In contrast with the preceding analysis, one should thus find the
dynamical symmetry group of the field theory with ${\cal A}^\prime$
to be U(2) having the compact group manifold 
${\cal M}^\prime({\rm U}(2)) \cong S^3$, 
as it is obvious from the defining quadratic form (\ref{H'}). 
However, the dynamical symmetry group is again U(1,1).
This result on continuous global symmetries hidden in (\ref{A}) 
and (\ref{A'}) looks somewhat enigmatic, if one takes into account 
only the equations of motion and ignores the corresponding discrete 
symmetries. Indeed, the action ${\cal A}$ is odd under the parity
and time-reversal transformations, 
$P,T: {\cal A} \rightarrow -{\cal A}$, 
although it describes physical states with the spins $-1$ and $+1$,
being thus respective to a $P,T$-invariant system. The price for such
a disorder is that the reduced phase space Hamiltonian $H^*$ is not 
positive-definite. On the other hand, ${\cal A}^\prime$ is parity 
and time-reversal invariant, and its reduced phase space Hamiltonian
$H^{\prime *}$ is positive-definite, but now the oscillator-like
operators have the following commutation relations:
\begin{equation}
[\widehat{z}_\epsilon(p),\widehat{\bar z}{}_{\epsilon^\prime}(q)]
= - \epsilon \delta_{\epsilon\epsilon^\prime} 
\delta(\vec{p} + \vec{q}).
\label{1'}
\end{equation}  
As a consequence, we get 
\begin{equation}
{(\widehat{z}_\epsilon)}^\dagger = 
- \epsilon \widehat{\bar z}{}_\epsilon
\label{2'}
\end{equation} 
for the Hermitian conjugation with respect to our positive-definite
scalar product $(\,,)$. The last two relations are essentially 
different from the corresponding eqs. (\ref{1}) and (\ref{2}). 
Taking into account these properties, we obtain the densities of the
dynamical symmetry group U(1,1) generators of the system (\ref{A'}),
(\ref{H'}) to be of the form
\begin{equation}
u^\prime = \frac{1}{2}\left( \widehat{\bar z}{}_-\widehat{z}_- 
+ \widehat{\bar z}{}_+\widehat{z}_+ \right), \quad
q^\prime_0 = \frac{1}{2}\left( \widehat{\bar z}{}_-\widehat{z}_- 
- \widehat{\bar z}{}_+\widehat{z}_+ + 1 \right), \quad
q^\prime_+ = i \widehat{z}_+\widehat{z}_-, \quad
q^\prime_- = i \widehat{\bar z}{}_-\widehat{\bar z}{}_+ .
\end{equation}
They satisfy the same relations as do the above described 
(non-primed) U(1,1) generators. 

Note that, analogous to (\ref{chi}) and (\ref{A}), the action 
${\cal A}^\prime$ is related to the constraint operator 
$\widehat\chi$ of the pseudoclassical model (\ref{LT}), 
only that averaged over the state space with the metric 
$\langle\, , \rangle$. Going back to the one-particle 
level of (\ref{A'}) and seeking for the new U(1,1) dynamical
symmetry group generators to be Hermitian with respect to the 
scalar product $\langle\, , \rangle$, one gets the desirable 
set of the physical operators by appropriate redefinitions of the 
U(1,1) quantum-mechanical generators $Q_\alpha$, $\alpha=0,1,2$, 
and $U$ from ref. \cite{NP1}. The new operators $Q^\prime_\alpha$ 
related to the system (\ref{A'}) form $su(1,1)$ symmetry and 
$s(2,1)$ supersymmetry \cite{CrRit} algebras, as it was the 
case of ref. \cite{NP1}.

We have observed that the hidden symmetries of the planar free
field systems revealed when explicitly modelling the underlying
spin dynamics at the one-particle level, appeared to be the
manifest (dynamical) symmetries of the corresponding field theories' 
reduced phase space Hamiltonian. And again, as well as for the 
dynamical picture of the pseudoclasical gauge systems \cite{NP1,NP2}, 
the discrete symmetries turned out to be of crucial importance for
the continuous global symmetries of these field theories. 
However, we must mention that it is rather difficult, if only 
feasible, to explain the hidden {\it supersymmetries} of the 
systems under consideration working on exclusively the field 
theory level, say, through some quirky fermionization. It seems 
very likely that in order to see the dynamical symmetries to be 
accompanied by the supersymmetries leading to non-standard 
super-extensions of the Poincar\'e group \cite{GPS,NP1}, it 
is necessary to investigate the corresponding pseudoclassical 
models.

\section{Concluding remarks}

We have analyzed hidden (dynamical) symmetries revealed in various
field configurations by means of the corresponding pseudoclassical
particle models. The motivation for the research into these field 
systems is usually based on the claim about their relevance to 
critical phenomena in planar physics. Therefore, the main problem 
to be further investigated in view of the present analysis is 
naturally to find any possible development of the discussed 
dynamical (super)symmetries in application to real physical 
processes. For this purpose, it worths to consider models 
with matter couplings to the free systems investigated here.

\small

\vskip5mm
\noindent{\bf Acknowledgement}
\vskip3mm
\noindent This work has been supported by the Alexander von Humboldt
Fellowship and the European Commission TMR Programme
ERBFMRX--CT96--0045 and ERBFMRX--CT96--0090.


\end{document}